\documentclass[twocolumn,10pt,pra,aps]{revtex4-1}
\def\by#1{#1,}
\def\and{and }
\def\yr#1{{(#1)}}
\def\paper#1{#1}
\def\jour#1{{\it #1}}

\def\vol#1{{\bf #1},}
\def\issue#1{}
\def\pages#1{\hbox{#1},}

\usepackage{amsmath}
\usepackage{amsfonts}
\usepackage{amssymb}
\usepackage{graphicx}
\usepackage{appendix}
\usepackage{placeins}

\newcommand{\pdiff}[3][]{\dfrac{\partial^{#1} #2}{\partial {#3}^{#1}}}
\newcommand{\dd}[1]{\ensuremath{\frac{d}{d{#1}}}}    % Total derivative in #1
\newcommand{\ddt}[1]{\dd t{#1}}    % Total time derivative in #1
\newcommand{\qp}[1]{\ensuremath{\!\left({#1}\right)}}
\newcommand{\qb}[1]{\ensuremath{\!\left[{#1}\right]}}

\begin{document}

\title{Planar hydraulic jumps in thin films: a regular solution against experiments}

\author{Alex V. Lukyanov}
\email{corresponding author,\\ a.lukyanov@reading.ac.uk}
\affiliation{School of Mathematical and Physical Sciences, University of Reading, Reading, RG6 6AX, UK}
\affiliation{P.N. Lebedev Physical Institute of the Russian Academy of Sciences, Moscow 119991, Russia}

\author{Tristan Pryer}
\affiliation{Department of Mathematical Sciences, University of Bath, Bath, BA2 7AY, UK}

\author{Edward Calver}
\affiliation{Department of Mathematical Sciences, University of Bath, Bath, BA2 7AY, UK}

\begin{abstract}
The formation of a planar hydraulic jump has been analysed in the framework of a full depth-averaged thin film model (DAM) with surface tension effects included. We have demonstrated regular weak solutions of the full DAM and analysed surface tension effects. It has been shown that surface tension effects within the parameter range relevant to the recent experiments are expected to be very weak and practically negligible. The developed methodology can be used in the analysis of laminar flow regimes and as a benchmark in developing full scale hydrodynamic models. 
\end{abstract}

\maketitle
 
\section{Introduction}

The phenomenon of a hydraulic jump, which is commonly observed in free surface flows, is well known to the research community. However, despite of almost a century of intensive research~\cite{Rayleigh1914, Tani1949, Watson1964, Craik1981, Bohr1993, Liu1993, Higuera1994, Higuera1997, Bohr1997, Avedisian2000, Bush2003, Ray2005, LowGravity2008, Das2008, Kasimov2008, Bonn2009, Ian2012, Ian2013, Limat2014, Ian2016, Ian2017, Ian2018, Benilov2019, Rohlfs2019, Bohr2019, Dhar2020, Linden2020, Vita2020, IanRaj2020}, the phenomenon itself and its mechanisms are still the subject of hot debates~\cite{Ian2017, Ian2018, Benilov2019, Rohlfs2019, Bohr2019, Dhar2020, Linden2020, Vita2020, IanRaj2020}. 

\begin{figure}[ht!]
\begin{center}
\includegraphics[trim=-0.5cm 2.cm 1cm 0.5cm,width=\columnwidth]{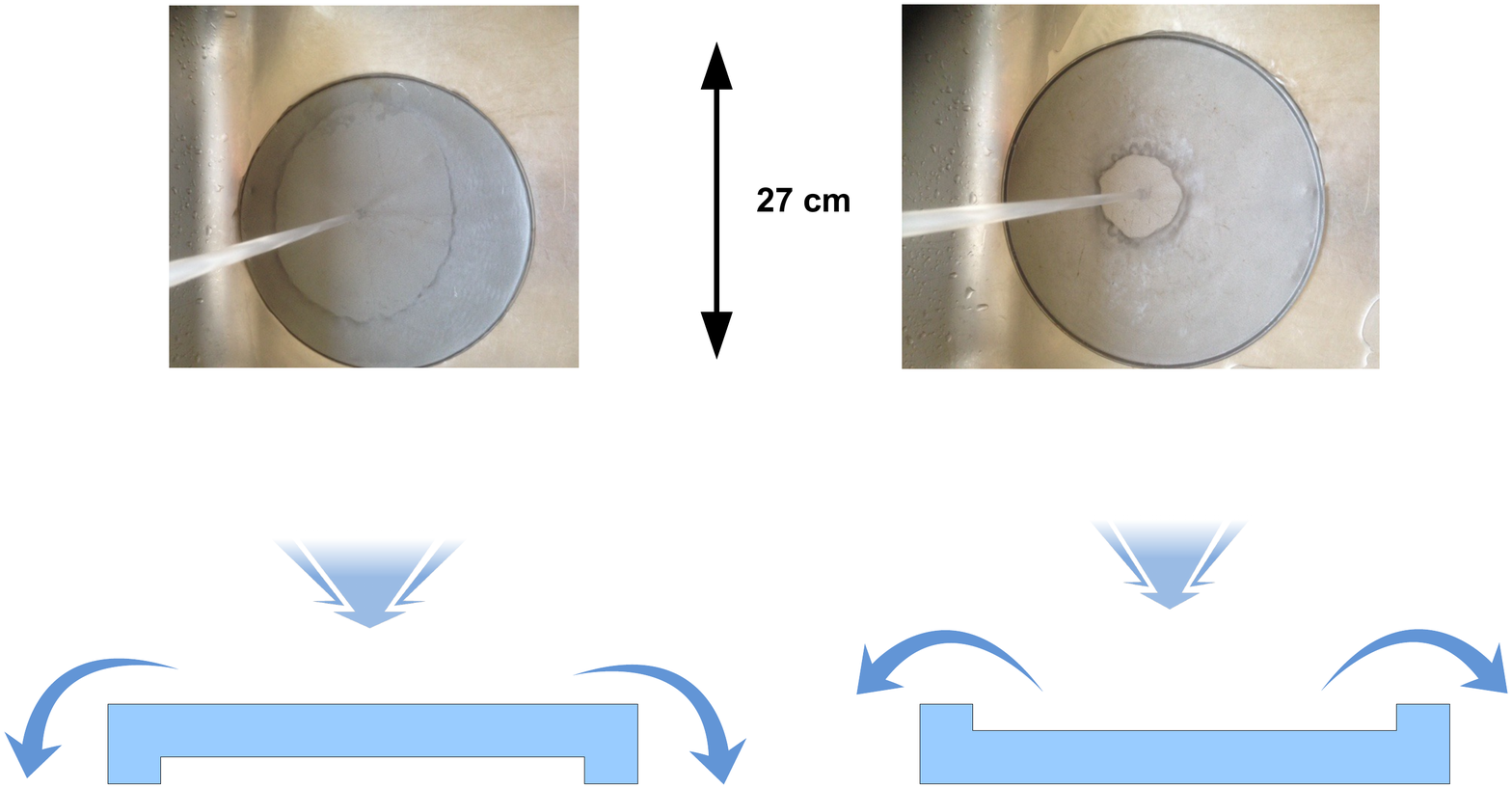}
\end{center}
\caption{Illustration of the circular hydraulic jump with different far-field conditions applied at a fixed value of the total flux in the jet $\approx 0.06\,\mbox{l/s}$: without (left) and with (right) the boundary edge. The boundary edge in the picture is about $\approx \, 3\mbox{mm}$. The substrate material is Teflon.} 
\label{Fig1}
\end{figure}

The phenomenon is quite interesting and to some extent intriguing, even though its appearance looks rather ordinary. The most commonly encountered are the circular hydraulic jumps, which can be observed practically in any kitchen using a tap and a lid, Fig. \ref{Fig1}. 

In general, hydraulic jumps could be of many different forms~\cite{Das2008, Bonn2009}, but the second most studied is a planar hydraulic jump observed in channel flows~\cite{Ray2005, Bonn2009, Ian2017, Dhar2020}. 

Recent debates have been instigated by new experimental evidence indicating that the position (the radius) of a circular hydraulic jump is practically independent of the gravity field direction to the substrate where the flow with a jump takes place~\cite{Ian2012, Ian2013, Ian2016, Ian2017, Ian2018}. That is when a liquid jet of a certain intensity impinges on solid walls oriented at different angles to the gravity force, the hydraulic jump radius measures almost the same at a fixed value of the total liquid flux in the jet. 

These observations in line with the previously reported data obtained in low gravity conditions~\cite{Avedisian2000, LowGravity2008} inspired a revision of the main mechanisms involved in the jump formation, emphasising the role of the surface tension, and the subsequent polemic~\cite{Ian2017, Ian2018, Benilov2019, Bohr2019, Rohlfs2019, Linden2020, IanRaj2020}. 

\begin{figure}[ht!]
\begin{center}
\includegraphics[trim=-0.5cm 3.3cm 1cm 0.5cm,width=\columnwidth]{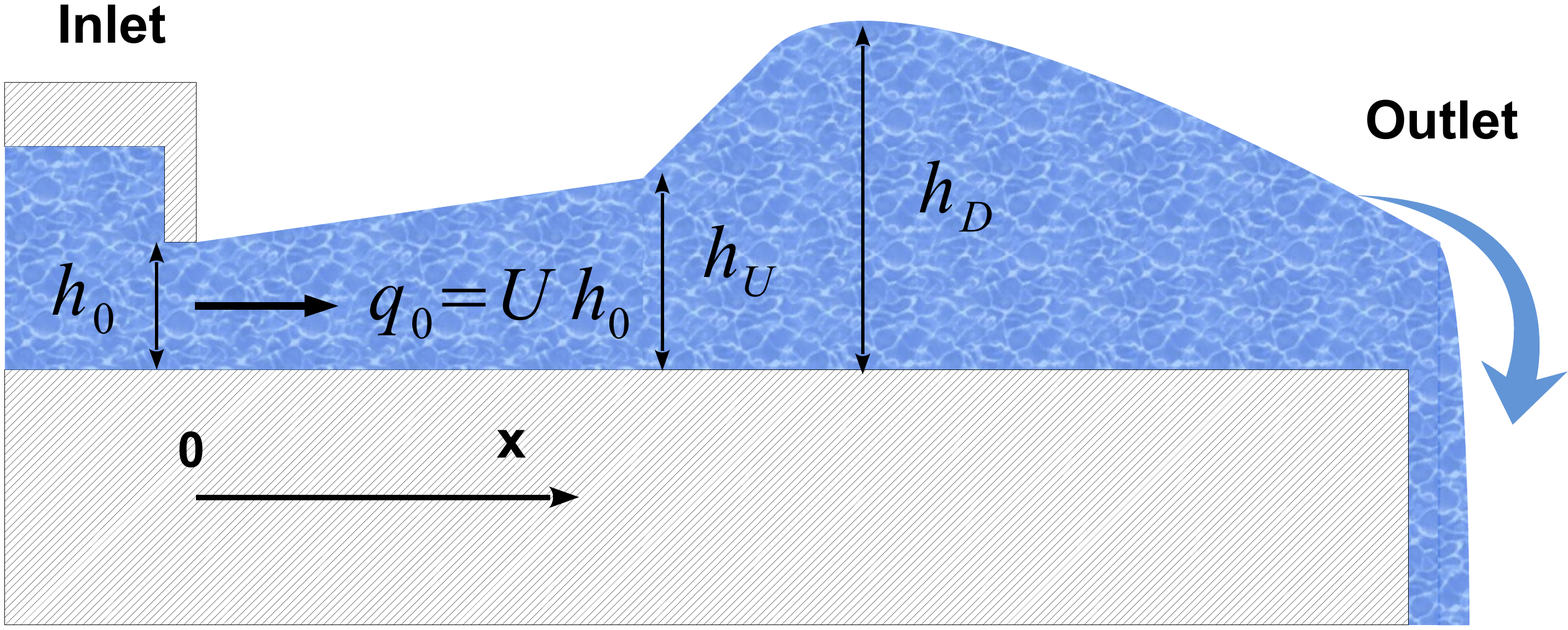}
\end{center}
\caption{Illustration of the planar hydraulic jump geometry.} 
\label{Fig3}
\end{figure}

Indeed, the role of gravity in the formation of a hydraulic jump is well known. Continuity of mass and momentum fluxes across the jump region treated as a discontinuity implies that at that point
\begin{equation}
\label{JCC}
\frac{Y\left(1 + Y   \right)}{2 Fr^2} + \frac{h_U}{R_0}Y We^{-1}=1,
\end{equation}
which can be used to estimate the position of the jump and its magnitude if somehow the free surface profiles are provided~\cite{Bush2003, Kasimov2008}. Here $Y=\frac{h_D}{h_U}$ is the ratio of the interface height $h_D$ in the downstream region, straight after the jump, to that in the upstream region $h_U$, just before the jump, Fig. \ref{Fig3}, $R_0$ is the jump radius, $Fr^2=\frac{6}{5}\frac{q_0^2}{g_0 h_U^3}$ and $We=\frac{6}{5}\frac{\rho q_0^2}{\gamma h_U}$ are the local values of the Froude and Weber numbers defined through the upstream height $h_U$, $q_0$ is the flux density per the unit length of the jump, $\gamma$ is surface tension and $g_0$ is the acceleration of gravity. The coefficient of $6/5$ has appeared due to a particular parabolic velocity profile used for averaging in this study, as in Appendix (\ref{KPA}).

As the aspect ratio of the layer height to the jump radius $R_0$ (in the case of a circular jump geometry), $\frac{h_U}{R_0}\ll 1$, is usually very small or zero in the case of a planar jump, contribution of the surface tension is often neglected leading to the classical B\'{e}langer equation~\cite{White2006} 
\begin{equation}
\label{JCCB}
Y=\frac{\sqrt{1+8Fr^2} - 1}{2}.
\end{equation}
This implies that the main mechanism of the hydraulic jump formation is supposed to be due to gravity. Somehow, the opposite was observed in the recent experiments~\cite{Ian2012, Ian2013, Ian2016, Ian2017, Ian2018}.  

In a simplified approach, the position of the jump can be roughly estimated as the critical point of the average velocity gradient~\cite{Kasimov2008} leading to a local condition
\begin{equation}
\label{JCFS}
Fr = 1.
\end{equation}
Note, criterion (\ref{JCFS}) is very approximate and can not in principle distinguish different far-field conditions, as it was rightly noticed in~\cite{Bohr2019}, while the far-field can dramatically affect the flow and the development of the hydraulic jump, as one can see from a simple experiment shown in Fig. \ref{Fig1}.

Based on the experimental observations~\cite{Ian2012, Ian2013, Ian2016, Ian2017, Ian2018}, the local criterion (\ref{JCFS}) was revisited and modified to~\cite{Ian2017, Ian2018} 
\begin{equation}
\label{STC}
We^{-1} + Fr^{-2} = 1,
\end{equation} 
which is supposed to be fulfilled in the case of an arbitrary jump geometry.

The modified criterion is still local and lacks information from the downstream conditions. But, it mitigates the effect of gravity while amplifying the role of the surface tension. As a result, other (different) trends are expected with the change of the controlling parameters of the flow (such as the total flux) and the liquid properties (such as surface tension)~\cite{Ian2018}. Condition (\ref{STC}) has been the subject of a polemic in the subsequent publications~\cite{Benilov2019, Rohlfs2019, Bohr2019, Dhar2020, Linden2020, Vita2020}.  

The authors of~\cite{Rohlfs2019}, using the full system of the Navier-Stokes equations, though in simplifying assumptions of a strictly laminar free-surface radial flow, have concluded that there are two typical flow conditions, capillary-dominant and gravity-dominant regimes, at any rate, the role of gravity cannot be eliminated or even substantially diminished. 

In a similar comparative study~\cite{Benilov2019}, the authors using the Navier-Stokes numerical solutions in a laminar flow regime with surface tension effects included, a shallow water approximation and a depth-averaged model (DAM), though neglecting surface tension, have also demonstrated the dominant role of gravity. At the same time, the laminar flow regime in full Navier-Stokes simulations was found to be unstable, when the coefficient of surface tension was exceeding some critical values. The critical values were well below the real values of the surface tension for such liquids as water. So that, the surface tension was found to be a destabilising factor.  

At the same time, in a recent publication~\cite{IanRaj2020}, the authors have demonstrated that there is compelling experimental evidence to doubt that the role of the surface tension is negligible, and the gravity dominates in the observed effects. 

One can summarise that at the moment the opinions are polarised that gravity still plays the dominant role. Moreover, no air-tight explanation for the effect of the invariance on the gravity force direction has been proposed. The role of surface tension, as a destabilising factor, also requires some further clarification to understand the extent that this mechanism can affect the formation of the transient region between the two separated zones, the upstream and the downstream regions. 

The purpose of the current study is manifold. First of all, we would like to clarify the role of surface tension in the formation of a hydraulic jump. This problem is intertwined with the short scale structure of the hydraulic jump region and the flow itself, that is with the contribution from the smaller scale eddy (turbulent) motion and capillary waves, and with the existing methodologies to model such flow regimes. 

Indeed, there are three main methodologies (approaches) commonly used in the analysis of free surface flows engaging different levels of approximation, and as a consequence a different number of simplifying assumptions. That is, flows with a jump region have been analysed based on the full system of the Navier-Stokes equations, using a shallow water (SW) approach reminiscent of the Prandtl boundary layer equations and applying the DAMs~\cite{Benilov2019}. 

The numerical analysis of the full Navier-Stokes model requires the least number of assumptions mostly related to the spatial resolution. But simulations are hindered by the high computational costs, especially if the flow regime develops instabilities and eventually a turbulent state. For these reasons, the use of the full model in practice has been always limited to a nearly laminar flow regime, when the spatial resolution and the time dependent features were rather limited~\cite{Rohlfs2019, Benilov2019}. 

The SW models appeal to the small aspect ratio of the two characteristic dimensions in the free surface flows when the liquid layer thickness is much smaller than the longitudinal length scales. The model can resolve some transient features in the flow and the lateral velocity profile with slow variations along the flow directions. But, the turbulent, short scale eddy motion, cannot be in principle resolved by the SW approach due to the approximation. Moreover, despite the simplifications, the SW approach still requires numerical tools to obtain a solution and is still can be computationally expensive to test in view of Courant-Friedrichs-Lewy stability restrictions.

The DAM approach introduces another level of averaging and approximation, and have the advantage of available analytical solutions and relatively low cost numerical solutions. The averaging over the layer thickness is applied resulting in a greater simplified system of governing equations (\ref{CM})-(\ref{ThinFilmII}) but requires closure in the form of a lateral velocity profile as in Appendix (\ref{ansatz}), for example. While the lateral velocity profiles required to define the coefficients in the model are to some extent assumed, the actual variations in the model parameters have been found non-essential~\cite{Vita2020}, as that any velocity profile still has to satisfy the boundary conditions. 

Therefore, the DAMs have been widely accepted as a reasonable approximation in many practical applications, but the contribution from the surface tension has been so far neglected resulting in discontinuous solutions. From the analytical solutions point of view, this was understandable, since surface tension contribution led to derivatives of the third order with no observable analytical solutions available. From the numerical simulations point of view, the progress was also surprisingly slow, which was also related to the high order derivatives. So that no continuous solutions have been developed so far.  

The common trend in the use of DAMs was to obtain discontinuous analytical solutions neglecting surface tension effects in the separated by the jump upstream and downstream regions and then to apply additional arguments, such as conditions (\ref{JCCB}) and (\ref{JCFS}) to identify the jump position. It is informative, that the layer averaging is still applied to obtain practical criteria such as (\ref{JCCB}) and (\ref{JCFS}). But the neglect of the surface tension effects left room for speculation as to the applicability of the DAMs (otherwise very efficient) somehow implying that the DAMs have always inherent singularities leading to discontinuous solutions, while the use of the next level models, such as the SW models has the advantage of continuity. 

So, the first purpose of this study is to develop reliable weak continuous solutions (and methodologies to obtain them) of the full system of DAM equations including the awkward third order terms responsible for the surface tension effects, to investigate the role of the surface tension in the formation of the jump region. The only disadvantage in comparison to the full Navier-Stokes model appears to be the resolution of the eddy motion. So that in the laminar flow regime, the weak solutions will be almost exact up to the minor corrections due to uncertainties with the lateral velocity profile.

Here, we would like to demonstrate, that the DAMs can always provide a regular solution, which carries, in fact, all the features available in the SW approximation. The comparative analysis of the three main approaches in~\cite{Benilov2019} has shown that the DAMs even in the presence of a critical point (due to the neglect of the surface tension terms, we argue) can catch the main features of the hydraulic jump effect. We will further stress and develop this point by demonstrating that the full DAM can provide a regular solution continuously linking the two regions of the flow and taking into account the effects of surface tension in full.

The second main task we are going to pursue is to analyse contribution from the small scale eddy motion. In the first approximation, we introduce effective eddy viscosity, as in~\cite{Bonn2009}, to account for the effect of the small scale motion on the averaged flow profiles.

The importance of effects of turbulence has been stressed in several studies involving planar and circular hydraulic jumps~\cite{Watson1964, Bonn2009, Ian2016}. The experimental study of planar hydraulic jumps in channels has demonstrated that eddy viscosity in both regions, the upstream and the downstream, exceeds the physical dynamic viscosity of the liquid by a factor of four. This means that the turbulence effects are by far more important than the average structure of the lateral velocity field producing only minor effects. 

In this flow regime, we will test, using the continuous solutions developed, how the averaging of the small scale flow structure affects the global solutions. We will demonstrate limitations of the full DAM and, at the same time, as a byproduct, develop means for a clear separation of different effects contributing to the formation of the free surface profile with a hydraulic jump.  The advantage of the DAM in this case is that it can clearly split up the effects of the surface tension and the eddy motion.

In the current study, we will concentrate on the planar hydraulic jumps, where effects of the finite jump radius are expected to be absent. We will develop weak solutions to the full system of the DAM equations including third order (spatial derivative) terms responsible for the surface tension effects. Using the DAM methodology, we will analyse parametric dependencies, including the effect of surface tension, and provide a comparative analysis of the experimental data in~\cite{Bonn2009}, where the turbulent effects manifest in full. We compare the obtained continuous solutions with the approximate theory and the criteria (\ref{JCCB}).  

In what follows, we first provide a mathematical model relevant to the DAM approximation and briefly revisit previous analytical results involving discontinuous solutions.

\begin{figure}[ht!]
\begin{center}
\includegraphics[trim=-0.5cm 2.cm 1cm 0.5cm,width=\columnwidth]{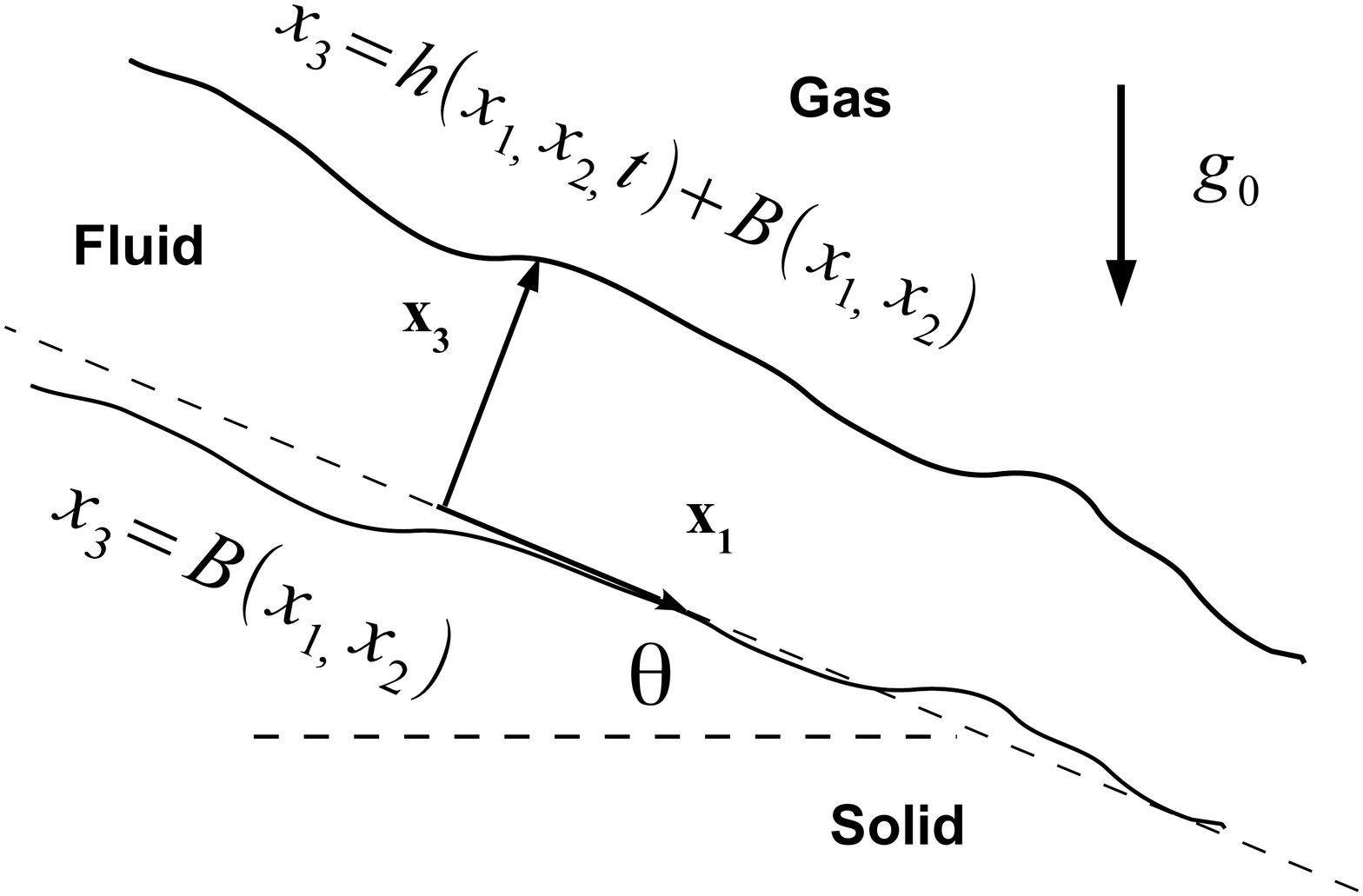}
\end{center}
\caption{Illustration of the thin film flow geometry.} 
\label{Fig2}
\end{figure}

\section{The mathematical model}

The mathematical model we utilise is relatively well understood. As such, we only sketch its derivation in the Appendix, while details can be found elsewhere~\cite{Matar2009}. The problem setup is illustrated schematically in Fig. \ref{Fig2}. This is a three-dimensional viscous flow of a Newtonian liquid at a solid wall located at $x_3=B(x_1,x_2)$ with a free surface parametrised accordingly $x_3=h(x_1,x_2,t) + B(x_1,x_2)$, Fig. \ref{Fig2}. 

The problem is characterised by two different length scales $H$ and $L$ in the vertical, $x_3$, and the horizontal $x_1, x_2$ directions respectively. Angle $\theta$ characterises the inclination of the flow plane to the direction of the gravity field $\bf g_0$, Fig. \ref{Fig2}.  In the thin film approximation taken in the study $H/L=\delta \ll 1$ is assumed to be a small parameter. 

The non-dimensional formulation is achieved by introducing reduced variables, that is the coordinates $x_1=\hat{x}_1/L$, $x_2=\hat{x}_2/L$, $x_3=\hat{x}_3/H$, velocities $v_1=\hat{v}_1/U$, $v_2=\hat{v}_2/U$, $v_3=\hat{v}_3/\delta U$, time $t/t_0$ and pressure $p=\hat{p}/p_0$. Here $U$ is characteristic velocity, $t_0=L/U$ is the timescale, $p_0=\rho U^2$ is the characteristic pressure in the inertial range, $\rho$ is the liquid density.  

The system of the governing equations in the thin film approximation is obtained by introducing averaged (over the layer) quantities 
$$
q_{1,2}=\int_{B}^{h+B}\,v_{1,2}\, dx_3.
$$ 

As a result of the averaging procedure, see details in Appendix and \cite{Matar2009},
\begin{equation}
\label{CM}
\frac{\partial h}{\partial t}  + \frac{\partial q_1}{\partial x_1} + \frac{\partial q_2}{\partial x_2} = 0,
\end{equation}

\begin{equation}
\label{ThinFilmI}
\frac{\partial q_1}{\partial t} + \frac{6}{5} \frac{\partial }{\partial x_1} \left( \frac{q_1^2}{h} \right) +  \frac{6}{5} \frac{\partial }{\partial x_2} \left( \frac{q_1q_2}{h} \right) = 
\end{equation}
$$
h\, \left\{ Ka_{\parallel}\sin\theta -   \pdiff{p}{x_1} \right\} -\frac{3}{Re}\frac{q_{1}}{h^2}
$$
and
\begin{equation}
\label{ThinFilmII}
 \frac{\partial q_2}{\partial t} + \frac{6}{5}\frac{\partial }{\partial x_1} \left( \frac{q_1q_2}{h}\right) +  \frac{6}{5} \frac{\partial }{\partial x_2} \left( \frac{q_2^2}{h}\right)=
\end{equation}
$$
-h\,  \pdiff{p}{x_2}  -\frac{3}{Re}\frac{q_{2}}{h^2},
$$
where pressure $p$ is given in the hydrostatic approximation by
\begin{equation}
\label{PrThinFilm}
p= p_a+Ka\, \cos\theta\, (h+B-x_3) -
\end{equation}
$$
\frac{1}{\hat{Ca}\, Re}\left(\frac{\partial^2 (h+B)}{\partial x_1^2}   +  \frac{\partial^2 (h+B)}{\partial x_2^2}\right). 
$$
Here $p_a$ is external gas pressure. The non-dimensional parameters of the problem are the Reynolds number, $Re=\delta \frac{\rho U H}{\mu}$, the Kapitza numbers, $Ka_{\parallel} =\frac{g_0 L}{U^2}$ and $Ka=\frac{g_0 H}{U^2}$ and $\hat{Ca} = Ca\, \delta^{-3}$, which is a renormalised Capillary number $Ca=\frac{\mu U}{\gamma}$, where $\gamma$ and $\mu$ are surface tension and dynamic viscosity of the liquid respectively.

To understand the functionality of the thin film system of equations (\ref{CM})-(\ref{PrThinFilm}) and the admissible solutions, we further simplify the problem to a one-dimensional case on a flat substrate, $B=0$, when a planar hydraulic jump is regularly observed. 

\subsection*{One-dimensional steady state problem}

In a one-dimensional case and on a flat substrate, $B=0$, 
\begin{equation}
\label{LE3}
\frac{\partial h}{\partial t}  + \frac{\partial q}{\partial x} = 0,
\end{equation}

\begin{equation}
\label{GE2D3sOne}
\frac{\partial q}{\partial t} + \frac{6}{5} \frac{\partial }{\partial x} \left( \frac{q^2}{h} \right) = - \frac{3}{Re}\frac{q}{h^2}  -   
\end{equation}
$$
h\left( Ka\cos\theta\, \pdiff{h}{x} -  Ka_{\parallel}\sin\theta - \frac{1}{\hat{Ca} Re} \pdiff[3]{h}{x}\right).
$$

In a steady state $q=const$, and  
\begin{equation}
\begin{split}
    \label{GE2D3sStSt}
0 &=
\frac{h^3}{\hat{Ca} Re} \pdiff[3]{h}{x} 
+ \pdiff{h}{x}\left(\frac{6}{5} q^2 - Ka\cos\theta h^3\right) 
\\ &\qquad +
h^3\, Ka_{\parallel}\sin\theta - \frac{3 q}{Re}.
\end{split}
\end{equation}

There are four non-dimensional parameters in the model reflecting the roles of gravity ($Ka$ and $Ka_{\parallel}$), surface tension ($\hat{Ca}$), viscosity and inertia ($Re$). The non-dimensional parameter $Ka$ is actually the inverse Froude number $Fr^2= \frac{6}{5}Ka^{-1}$. In what follows, we focus on the case $\theta=0$ (unless otherwise specified), that is when the gravity is perpendicular to the liquid layer.

\section*{Hydraulic jump as a discontinuity}

A set of typical experimental results in planar channel flows when a hydraulic jump was commonly observed in both laminar (inset) and turbulent flow regimes are shown in  Fig. \ref{Fig4} with the parameters summarised in Table \ref{Table1}. The main input and output parameters are the flux density $q_0$, initial film thickness $h_0$ at the entrance of the flow and the position of the jump $X$ with respect to the entry point. In this study, we define the jump position as the inflection point of the free surface profile. 

As one can see, the dimensional flux density in the experiments $q_0$ spans 
$$
7\cdot 10^{-5}\, \mbox{m}^2/\mbox{s} \le q_0 \le 2\cdot 10^{-3} \, \mbox{m}^2/\mbox{s}.
$$   
The range of the non-dimensional parameters of the problem based on the upstream conditions then suggests a set of potentially asymptotic parameters $Ka\ll 1$, $\hat{Ca}\gg 1$ and $Re\gg 1$, if we define $H=h_0$ and the characteristic velocity by means of $q_0 = U h_0$, so that non-dimensional $q=1$. The horizontal length scale $L$ is chosen on the basis of $h_0$ by setting $\delta=0.1$. That is in the case of water
$8 \le Re\le 200$, $4 \le \hat{Ca}\le 140$ and $3\cdot 10^{-4}\le Ka \le 4\cdot 10^{-2}$. 

If contribution of the surface tension, the third order differential operator, is ignored away from the transition region at $\hat{Ca}Re \gg 1$, equation (\ref{GE2D3sStSt}) is reduced to
\begin{equation}
\label{GE2D3sStStSimp}
\pdiff{h}{x}\left\{ \frac{6}{5} q^2 - Ka\, h^3\right\} =\frac{3q}{Re}.
\end{equation}

Note, in our choice of non-dimensional parameters ($H=h_0$), $q=1$ and the equation can be further simplified. In an arbitrary normalisation, one can re-scale the non-dimensional variables $x,h$ according to
\begin{equation}
\label{SCtp}
x=\tilde{x} \frac{Re}{3}\left(\frac{6}{5}\right)^{4/3}\frac{q^{5/3}}{{Ka}^{1/3}} ,\quad h=\tilde{h} \left(\frac{6}{5}\right)^{1/3}\frac{q^{2/3}}{{Ka}^{1/3}}
\end{equation}
to bring (\ref{GE2D3sStStSimp}) into a parameter-free equation
\begin{equation}
\label{GE2D3sStStSimpPL}
\pdiff{\tilde{h}}{\tilde{x}}\left\{ 1 - \tilde{h}^3\right\} =1.
\end{equation}
The scaling (\ref{SCtp}) suggests that the position of the hydraulic jump should be proportional to $X\propto q_0^{5/3}$, which is indeed observed in the experiments~\cite{Ray2005, Bonn2009, Ian2017, Dhar2020}, as one can see from Fig. \ref{Fig4}, linear fits to the data. 

Solving equation (\ref{GE2D3sStStSimp}) with $h(0)=1$ and $q=1$, one gets
\begin{equation}
\label{Branch1}
\frac{6}{5}(h-1) - \frac{Ka}{4} (h^4-1) = \frac{3}{Re} x.
\end{equation}
The obtained solution has a critical point at 
$
h_c^3=\frac{6}{5\, Ka}
$
and covers the range
$$
x\le x_c = \frac{6}{5}(h_c-1) - \frac{Ka}{4} (h_c^4-1),
$$
see Fig. \ref{Fig5}. The obtained solution can not connect the upstream region to the downstream far-field so the discontinuity is inherent to this approximation due to the low order of the differential equation with a critical point. 

The second branch describing the downstream free surface profile, Fig. \ref{Fig5}, is obtained from the far-field condition at $x=l_s$ in a similar way as in~\cite{Kasimov2008, Bonn2009} by placing the critical point at $x=l_s$, that is $h=h_c$ at $x=l_s$,
and the the second branch
\begin{equation}
\label{Branch2}
\frac{6}{5}(h-h_c) - \frac{Ka}{4} (h^4-h_c^4) = \frac{3}{Re} (x-l_s).
\end{equation}

The obtained discontinuous solution can not inform us about the position of the jump, which may occur anywhere $x<x_c$, subject to an additional condition of the mass and momentum flux continuity~\cite{Bush2003, Kasimov2008}.

The discontinuous solution can not, of course, describe the shape of the jump itself, but, despite the neglect of the surface tension term, can match observations away from the jump region in a laminar case.

\subsection*{Laminar flow regime}
The formation of a planar hydraulic jump in a laminar flow regime ($8\le Re\le 16$) has been recently studied in detail with free surface profiles observed at different inclination angles $-0.6^{\circ}\le \theta\le 1.5^{\circ}$~\cite{Dhar2020}. In a particular case of $\theta=0^{\circ}$, the profile is demonstrated in Fig. \ref{Fig41}, where the branches of the analytical solutions (\ref{Branch1}) and (\ref{Branch2}) are shown for comparison. As one can observe, the solutions match the experimental profiles very well away from the jump region. In particular, one can see the typical linear dependence in the upstream part as is expected from (\ref{Branch1}). The location of the hydraulic jump according to (\ref{JCCB}) is practically at the point where the experimental profile starts to deviate from the upstream branch of the solution. The subsequent values of the local non-dimensional parameters at that point are $Fr^2\approx 7$ and $We\approx 0.4$ ($0 < x < x_c$, \, $1 < Fr^2 < 120, \,\, 0.2 < We < 1$), so that neither (\ref{JCFS}) nor (\ref{STC}) are fulfilled at the point of the jump or at the inflection points. 

The flow rate dependence of the jump position $X$ in the laminar case follows the well-known trend $X\propto q_0^{5/3}$, as well as its dependence on the inclination angle, Fig. \ref{Fig4}.

\subsection*{Turbulent flow regime}

We consider now particular experimental observations in water flows~\cite{Bonn2009}, where detailed measurements of the surface profiles have been conducted in a turbulent flow regime ($Re\approx 100$), Fig. \ref{Fig6}. The free surface profile has again that distinctive, almost linear upstream part, but the linear fit (\ref{Branch1}) only matches the profile, if the effective viscosity $\mu_{\epsilon}$ is about four times larger than that of water $\mu$ used in the experiments, leading to much lower effective Reynolds numbers. As it has been discussed in~\cite{Bonn2009}, the observed upstream free surface (linear) profile can be explained by the appearance of the eddy viscosity due to the flow turbulisation in the narrow channel. 

One can also observe that the turbulence contribution was non-uniform over the flow domain, as the downstream profile (\ref{Branch2}) using the same value of $\mu_{\epsilon}=const$ did not match the experimental observations. 

Therefore, additional terms associated with the turbulence were empirically introduced into a thin film model based on the mixing-length theory of Prandtl~\cite{White2006, Bonn2009} with the main assumption that the eddy viscosity $\mu_{\epsilon}$ is proportional to the value of the flux density $\mu_{\epsilon}\approx \rho k_{\epsilon}^2 q_0$ with $k_{\epsilon}=const$.  Such an assumption was supported by the observation that the inclination of the upstream free surface profile was practically independent of $q_0$, which was the feature absent in the laminar regime~\cite{Dhar2020}. 

Indeed, if we consider only the linear part of (\ref{GE2D3sStStSimp}), then $\frac{d h}{d x}\propto Re^{-1}\propto \mu_{\epsilon}\, q_0^{-1}$, so that if $\mu_{\epsilon}\propto q_0$, then the observed profile should show no dependence on the flux density value, as is indeed observed. The estimated values of $k_{\epsilon}$ were found at $k_{\epsilon}\approx 0.065$ leading to $\mu_{\epsilon}\approx 4.2\cdot 10^{-3}\,\mbox{Pa}\cdot\mbox{s}$ for water at $q_0=10^{-3}\,\mbox{m}^2/\mbox{s}$. 

\begin{figure}[ht!]
\begin{center}
\includegraphics[trim=-0.5cm 1.3cm 1cm 0.5cm,width=\columnwidth]{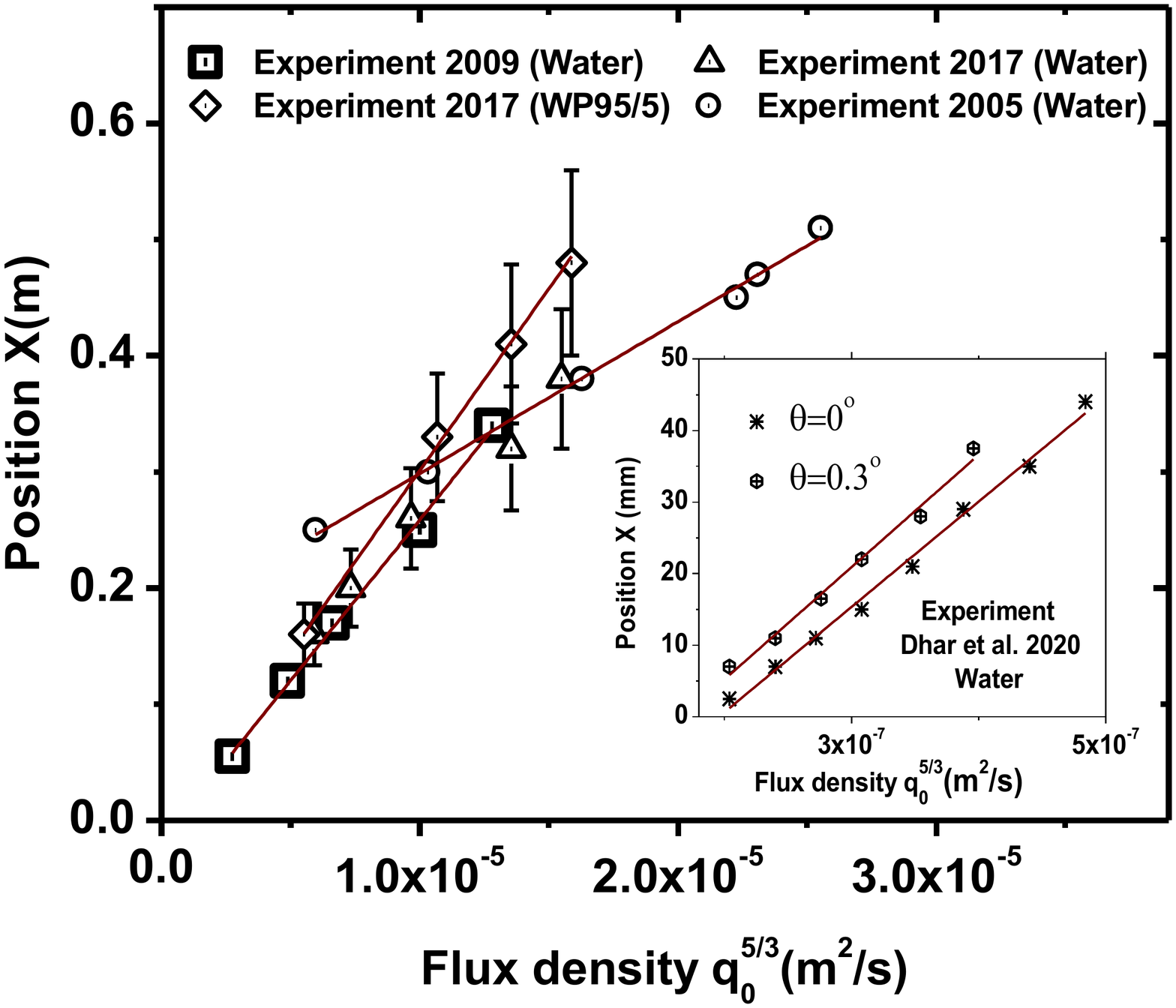}
\end{center}
\caption{Hydraulic jump position $X$ (symbols) as a function of the flow rate $q_0^{5/3}$ as is observed in the experiments~\cite{Ray2005, Bonn2009, Ian2017}, see Table \ref{Table1} for details. The insert shows experimental observations in the laminar case from~\cite{Dhar2020} at different inclination angles $\theta=0^{\circ}$ and $\theta=0.3^{\circ}$. The solid lines (brown) are linear fits $X\propto q_0^{5/3}$ to the experimental data.} 
\label{Fig4}
\end{figure}

\begin{figure}[ht!]
\begin{center}
\includegraphics[trim=0.5cm 1.3cm 1cm 0.5cm,width=\columnwidth]{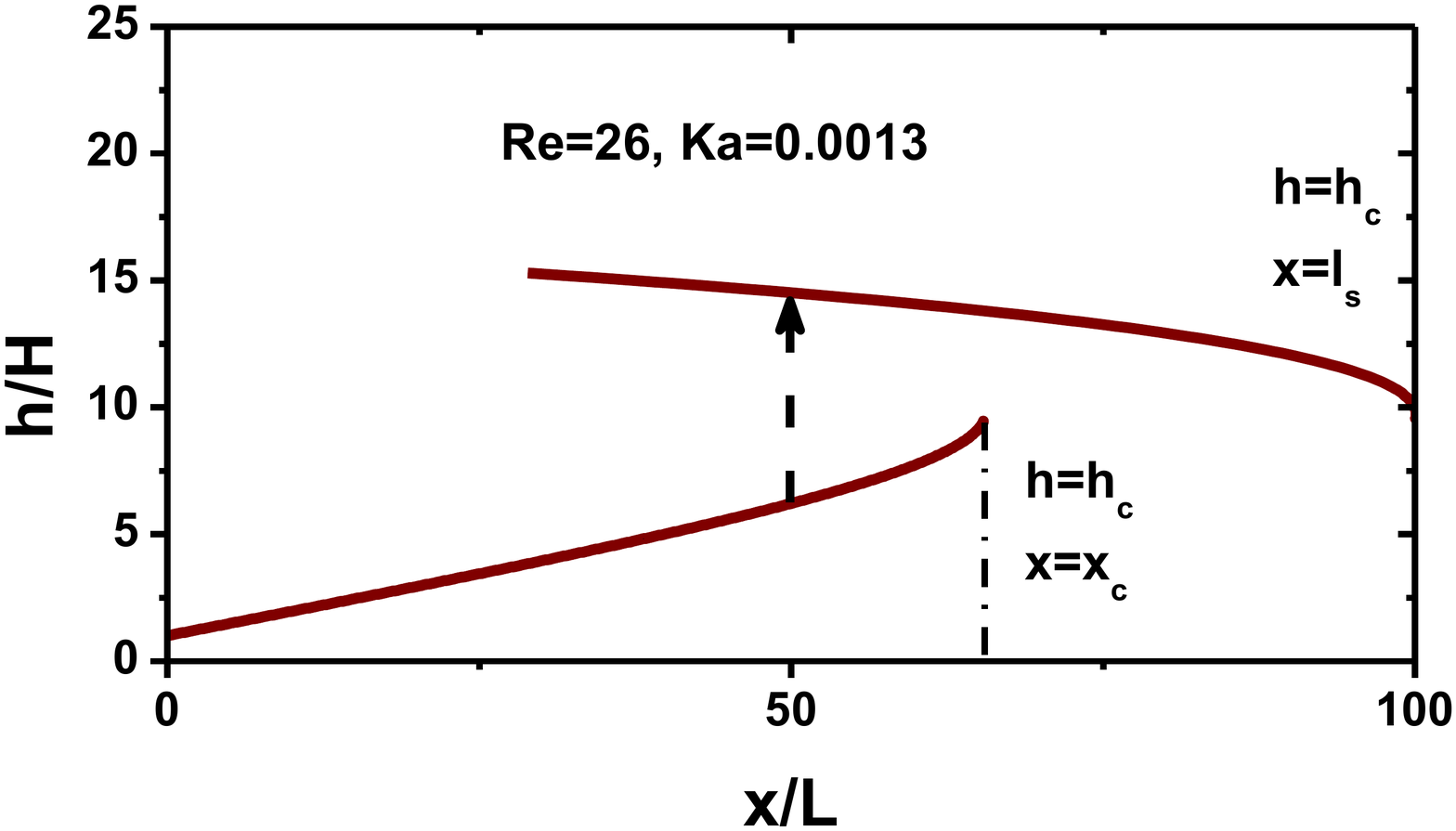}
\end{center}
\caption{Illustration of the two branches of the steady state solutions, (\ref{Branch1}) and (\ref{Branch2}), at $Re=26$, $Ka=0.0013$, $\hat{Ca}=108$ and $l_s=100$. The arrow is the jump position according to (\ref{JCCB}).} 
\label{Fig5}
\end{figure}

\begin{figure}[ht!]
\begin{center}
\includegraphics[trim=0.5cm 4.3cm 1cm 0.5cm,width=\columnwidth]{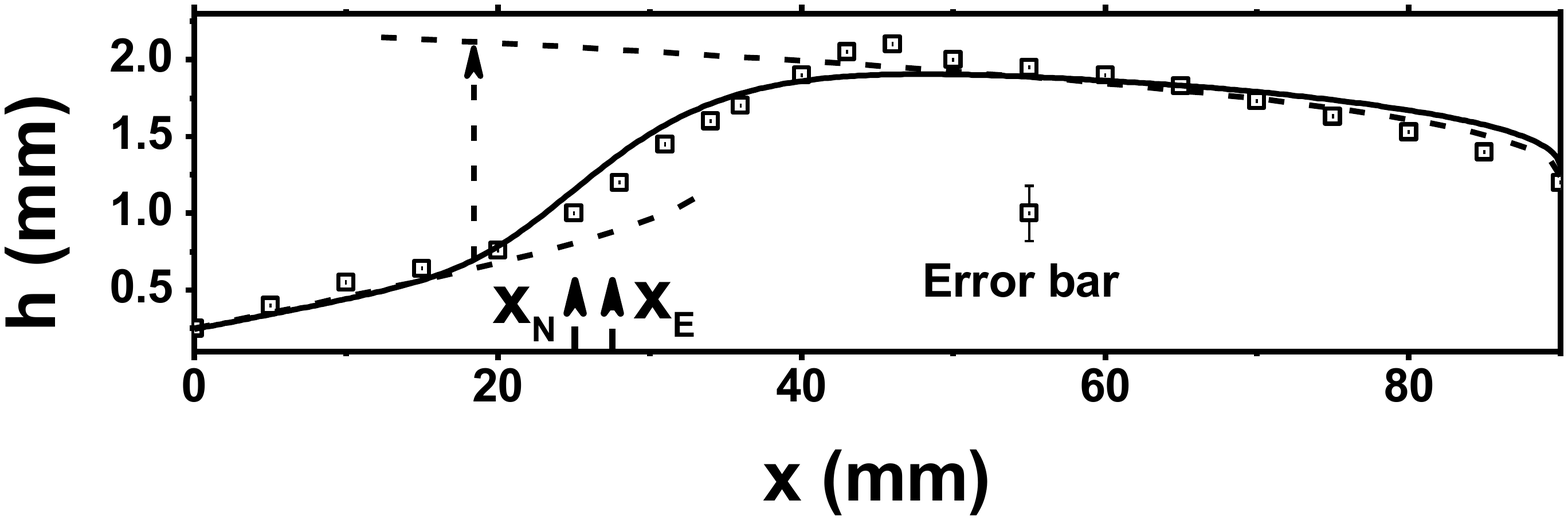}
\end{center}
\caption{Illustration of the experimental free surface profile (symbols) in a laminar regime from~\cite{Dhar2020} at $\theta=0^{\circ}$, $q_0=1.25\cdot 10^{-4}\,\mbox{m}^2/\mbox{s}$, $Re=12.5$, $\hat{Ca}=6.9$ and $Ka=0.0098$. The dashed lines are analytical solutions (\ref{Branch1}) and (\ref{Branch2}), the solid line is the numerical solution of (\ref{LE3}) - \eqref{GE2D3sOne}. The arrows are the positions of the inflection points of the experimental and numerical profiles, $x_E$ and $x_N$, respectively, and the jump according to (\ref{JCCB}).} 
\label{Fig41}
\end{figure}

\begin{figure}[ht!]
\begin{center}
\includegraphics[trim=0.5cm 1.3cm 1cm 0.5cm,width=\columnwidth]{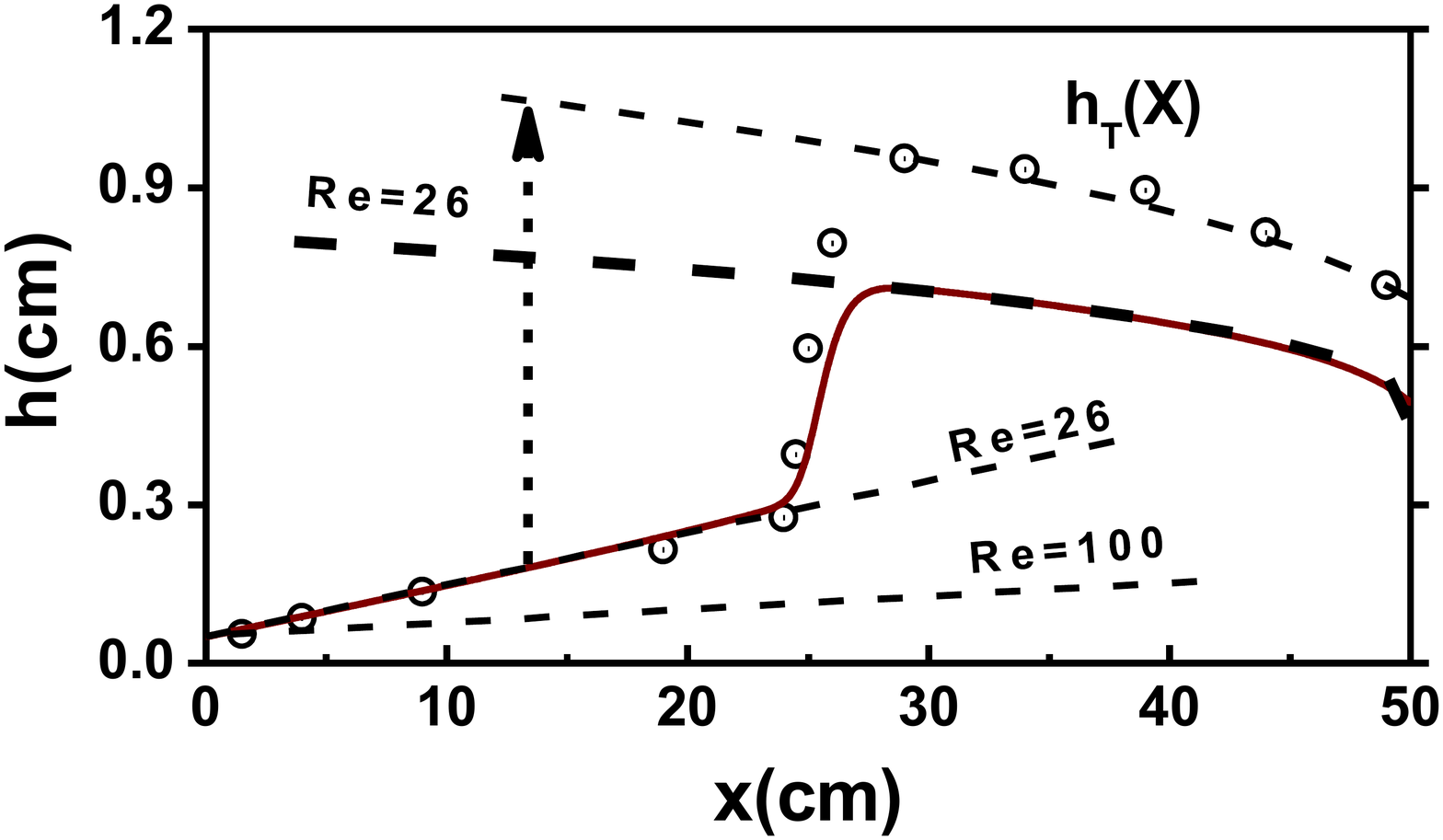}
\end{center}
\caption{Experimental (symbols, \cite{Bonn2009}) and numerical (solid line, brown, solution to (\ref{LE3})-(\ref{GE2D3sOne})) free surface profiles in water channel flows at $q_0=10\,\mbox{cm}^2/\mbox{s}$, at $\theta=0^{\circ}$, $Re=26$, $\hat{Ca}=108$ and $Ka=1.3\cdot10^{-3}$. The dashed lines are the linearised solutions (\ref{Branch1}) in the upstream at $Re=26$ and $Re=100$ respectively, and the downstream solution (\ref{Branch2}) at $Re=26$. The upper dashed line $h_T(x)$ is the inverse of (\ref{STB2}) at $a_{\epsilon}=1$. The dashed arrow is positioned at the jump according to (\ref{JCCB}).} 
\label{Fig6}
\end{figure}

The appearance of the turbulent motion was attributed to a relatively narrow channel used in the experiments, Table \ref{Table1}. One should note though, as the authors did as well, that similar enhanced elevation of the upstream linear profile was observed in~\cite{Ray2005} with a much wider channel at comparable values of the other parameters. This was an indication that turbulent flow regimes were inherent to all experiments shown in Fig. \ref{Fig4}.

There were two different branches in the modified formulation~\cite{Bonn2009}. They are given by, using scaling (\ref{SCtp}) with $\mu=\mu_{\epsilon}$,
\begin{equation}
\label{TB1}
\frac{d \tilde{h}}{d \tilde{x}} (1-\tilde{h}^3) = 1 + a_{\epsilon} \tilde{h}^2
\end{equation}
in the upstream region and by
\begin{equation}
\label{TB2}
\frac{d \tilde{h}}{d \tilde{x}} (1-\tilde{h}^3) \tilde{h} = 1 + a_{\epsilon} \tilde{h}^2
\end{equation}
in the downstream, where parameter $a_{\epsilon}\sim O(1)$ characterizes the velocity profile and the aspect ratio of the channel flow $h_c/d$~\cite{Bonn2009}. 

General solutions to (\ref{TB1}) and (\ref{TB2})
are given by
\begin{equation}
\label{STB1}
\tilde{x} = \frac{1}{2a_{\epsilon}^2}\ln(1 + a_{\epsilon} \tilde{h}^2) + \frac{1}{a_{\epsilon}^{1/2}}\arctan(a_{\epsilon}^{1/2} \tilde{h}) - \frac{\tilde{h}^2}{2a_{\epsilon}} + C
\end{equation}
and
\begin{equation}
\label{STB2}
\tilde{x} = \frac{1}{2a_{\epsilon}^2}\ln(1 + a_{\epsilon} \tilde{h}^2) - a_{\epsilon}^{5/2}\arctan(a_{\epsilon}^{1/2} \tilde{h}) + \frac{\tilde{h}}{a_{\epsilon}^2} - \frac{\tilde{h}^3}{3a_{\epsilon}} +C
\end{equation}
respectively. In the limit of $a_{\epsilon}\to 0$ corresponding to a wide channel, both solutions converge to (\ref{Branch1}).  

As one can see, the surface tension contribution was still neglected in the modified formulation, but as a result, the authors were able to obtain quite treatable analytic solutions to get a good comparison in some cases with their experiments. 

As far as the jump conditions (\ref{JCCB}) and (\ref{STC}) are concerned, none of them are fulfilled as one can observe in Fig. \ref{Fig6}, where the location of condition (\ref{JCCB}) is shown by an arrow, while at the transition point $x\approx 24\,\mbox{cm}$, the second condition  $Fr^{-2} + We^{-1}\approx 0.4$, so that it can only be fulfilled further downstream, though can not be ruled out. 

To note, the empirical mixing-length theory approach is ad hoc and discontinuous leading to different models applied in the upstream and downstream regions. As a result, it can not be directly implemented into the full DAM. To circumvent this limitation, we only used one element of that treatment in comparison with observation in turbulent flows, the modified, eddy viscosity $\mu_{\epsilon}$.  

In the next part, we will briefly discuss the numerical technique we utilised, and demonstrate regular solutions to the full DAM, (\ref{LE3}) - (\ref{GE2D3sOne}), with the surface tension term included. We study their parametric dependencies and attempt to model the liquid flows observed in~\cite{Bonn2009, Dhar2020}. 

\begin{figure}[ht!]
\begin{center}
\includegraphics[trim=-0.5cm 1.3cm 1cm -0.5cm,width=\columnwidth]{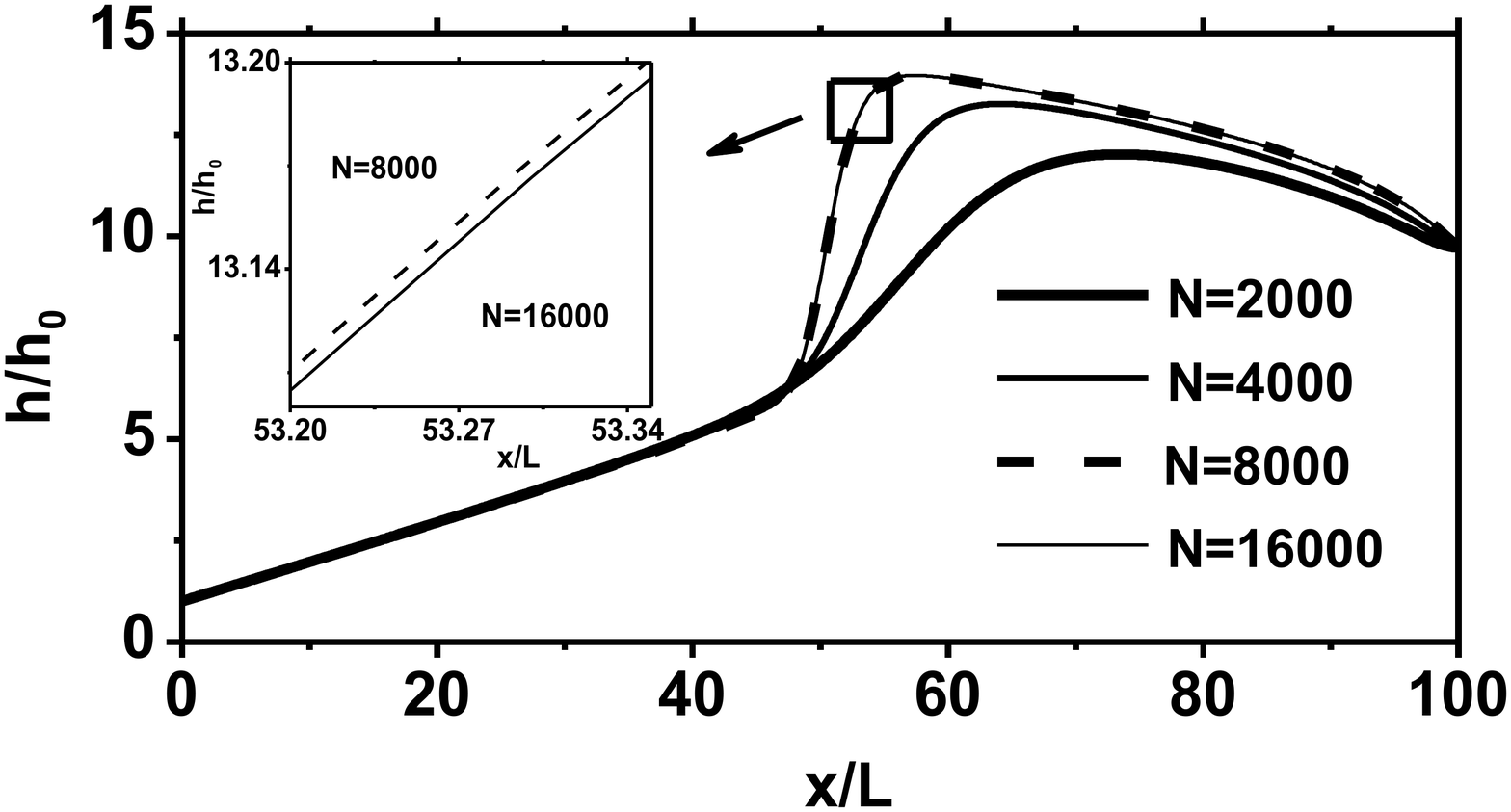}
\end{center}
\caption{Free surface profiles calculated numerically at $\theta=0^{\circ}$, $Re=26$, $\hat{Ca}=108$ and $Ka=0.0013$ and at different spatial resolutions $\Delta x=100/N$, where $N$ is the number of intervals.
\label{fig:num}} 
\end{figure}

\begin{figure}[ht!]
\begin{center}
\includegraphics[trim=0.5cm 1.3cm 1cm -0.5cm,width=\columnwidth]{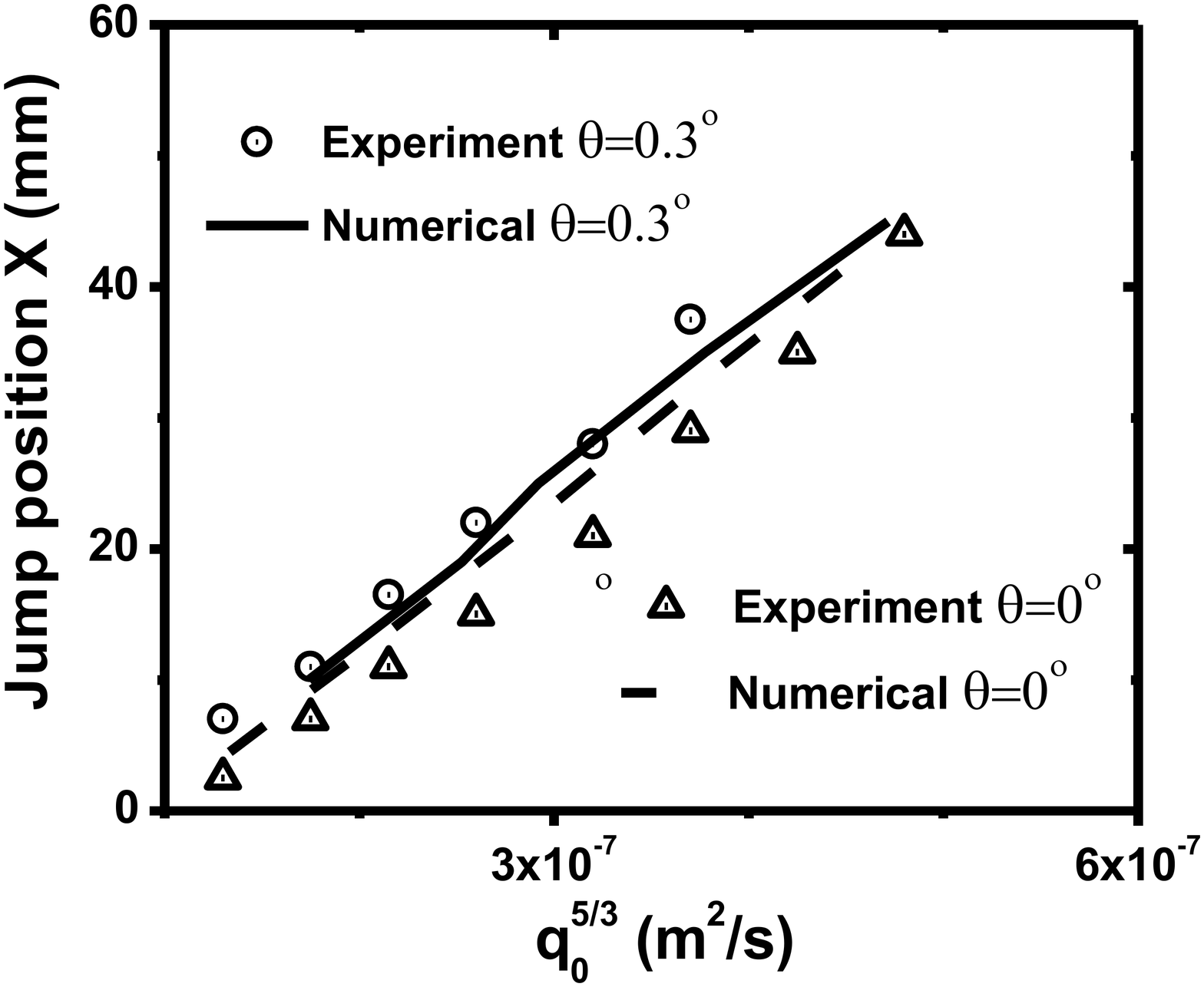}
\end{center}
\caption{Hydraulic jump position $X$ as a function of the flow rate $q_0^{5/3}$ in the experiments~\cite{Dhar2020} (symbols) and numerical simulations at $\theta=0^{\circ}$ (the dashed line) and at $\theta=0.3^{\circ}$ (the solid line).
\label{Fig81}} 
\end{figure}

\begin{figure}[ht!]
\begin{center}
\includegraphics[trim=0.5cm 1.3cm 1cm -0.5cm,width=\columnwidth]{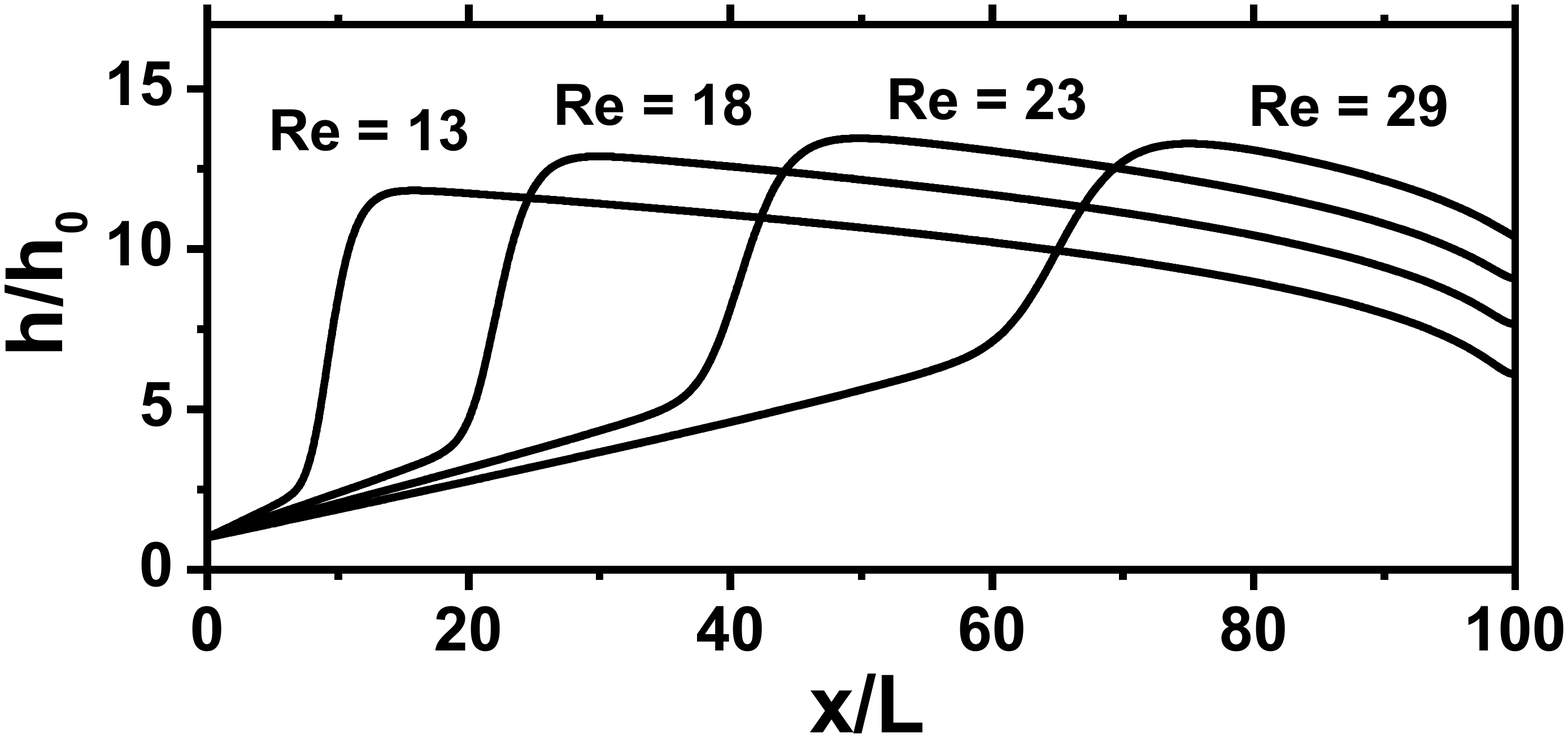}
\end{center}
\caption{Free surface profiles calculated numerically in the parameter range relevant to the experiments~\cite{Bonn2009} at $\theta=0^{\circ}$. The non-dimensional parameters were scaled up and down from the reference values $Re^{(0)}=26$, $\hat{Ca}^{(0)}=108$ and $Ka^{(0)}=1.3\cdot 10^{-3}$ to mimic variations of the flux according to $Re=Sc\, Re^{(0)} $, $\hat{Ca} = Sc \, \hat{Ca}^{(0)} $ and $Ka = Ka^{(0)}/Sc^2$, where $0.5\le Sc\le 1.1$.} 
\label{Fig7}
\end{figure}

\section*{Numerical discretisation and benchmarking}

We discretise (\ref{LE3}) - \eqref{GE2D3sOne} through a method of  lines approach utilising a finite volume spatial discretisation with a Lax-Friedrichs flux type. 

Let $\mathcal{T} = \{ x_i \}$ be a partition of the domain $[0,L]$ into cells $K_i = (x_{i-1/2}, x_{i+1/2})$. Here $x_i = \tfrac 1 2 (x_{i+1/2}+x_{i-1/2})$ denotes the midpoint of a cell $K_i$. Let $\Delta x_i = x_{i+1/2} - x_{i-1/2}$ denote the length of the cell. We only consider the case $\Delta x_i \equiv \Delta x$ for all $i$, however note that various adaptive strategies exist for this class of problems \cite{Pryer2015} that may be able to provide better resolution at the jump interface.

For exposition, we reformulate (\ref{LE3}) - \eqref{GE2D3sOne} in a conservative form
\begin{equation}
    \begin{split}
      \frac{\partial h}{\partial t}  + \frac{\partial q}{\partial x} = 0,
      \\
      \frac{\partial q}{\partial t}
      +
      \frac{\partial }{\partial x} F(q,h)
      -
      \frac{\partial }{\partial x} G(h)
      = - \frac{3}{Re}\frac{q}{h^2},
    \end{split}
\end{equation}
where
\begin{equation}
  \begin{split}
    F(q,h) &= \frac 6 5 \frac{q^2}{h} - \frac{Ka}{2} h^2
    \\
    G(h)
    &=
    \frac{1}{\hat{Ca} Re} \qp{\frac 12 \pdiff[2]{h^2} x - \frac 32 \qp{\pdiff h x}^2}.
  \end{split}
\end{equation}
Let $\chi_{K_i}$ denote the indicator function over the cell $K_i$, we then define our 
 numerical approximation
\begin{equation}
  \begin{split}
    H(x,t) &= \sum_i H_i(t) \chi_{K_i}(x)
    \\
    Q(x,t) &= \sum_i Q_i(t) \chi_{K_i}(x),
  \end{split}
\end{equation}
where $H_i, Q_i$ solve the following system of ODEs:
\begin{equation}
  \begin{split}
    \ddt H_i + \frac 1 {\Delta x} \qb{Q_{i+1/2} - Q_{i-1/2}} &= 0
    \\
    \ddt Q_i + \frac 1 {\Delta x} \qb{\mathcal F_{i+1/2} - \mathcal F_{i-1/2}}
    &+ \frac 1 {\Delta x} \qb{\mathcal G_{i+1/2} - \mathcal G_{i-1/2}}
    \\
    &\qquad = -\frac 3 {Re} \frac{Q_i}{H_i^2}
  \end{split}
\end{equation}
and $\mathcal F, \mathcal G$ represent approximations to $F$ and $G$
respectively. For our experiments we chose a Lax-Friedrichs flux
\begin{equation}
  \begin{split}
    \mathcal F_{i+1/2}
    &=
    \frac 1 2 \qp{F(U_i) + F(U_{i+1})} \\
    &\qquad - \max(\nabla F(U_i), \nabla F(U_{i+1})) \cdot \qp{U_{i+1} - U_i}
    \\
    \mathcal G_{i+1/2}
    &=
    \frac{W_{i+1} + W_i}{2},
  \end{split}
\end{equation}
where $\nabla F$ is the vector valued gradient of $F$, $U_i = (Q_i, H_i)$ and $W_i$ represents a standard central approximation to $G$.
This is formally a first order scheme in space, note that higher order
schemes are available including MUSCL and WENO schemes.

For the temporal discretisation, we use a third-order strong stability
preserving scheme. To ensure the method remains stable we make use of
an adaptive time-step chosen to ensure the Courant–Friedrichs–Lewy
condition is always met.

To test the method converges we fix parameters $Re = 26, \hat{Ca} = 108, Ka
= 0.0013$ and simulate solutions over a family of mesh sizes until a
steady state is found. We select $\Delta x \in [0.00625, 0.05]$ and plot specific
numerical profiles in Figure \ref{fig:num}. Notice that the method is quite diffuse for coarse mesh-scale and the position and profile of the jump is mesh-dependent. To ensure physically accurate results, all our experiments henceforth were obtained using $\Delta x = 0.00625$.

\begin{figure}[ht!]
\begin{center}
\includegraphics[trim=0.5cm 1.3cm 1cm -0.5cm,width=\columnwidth]{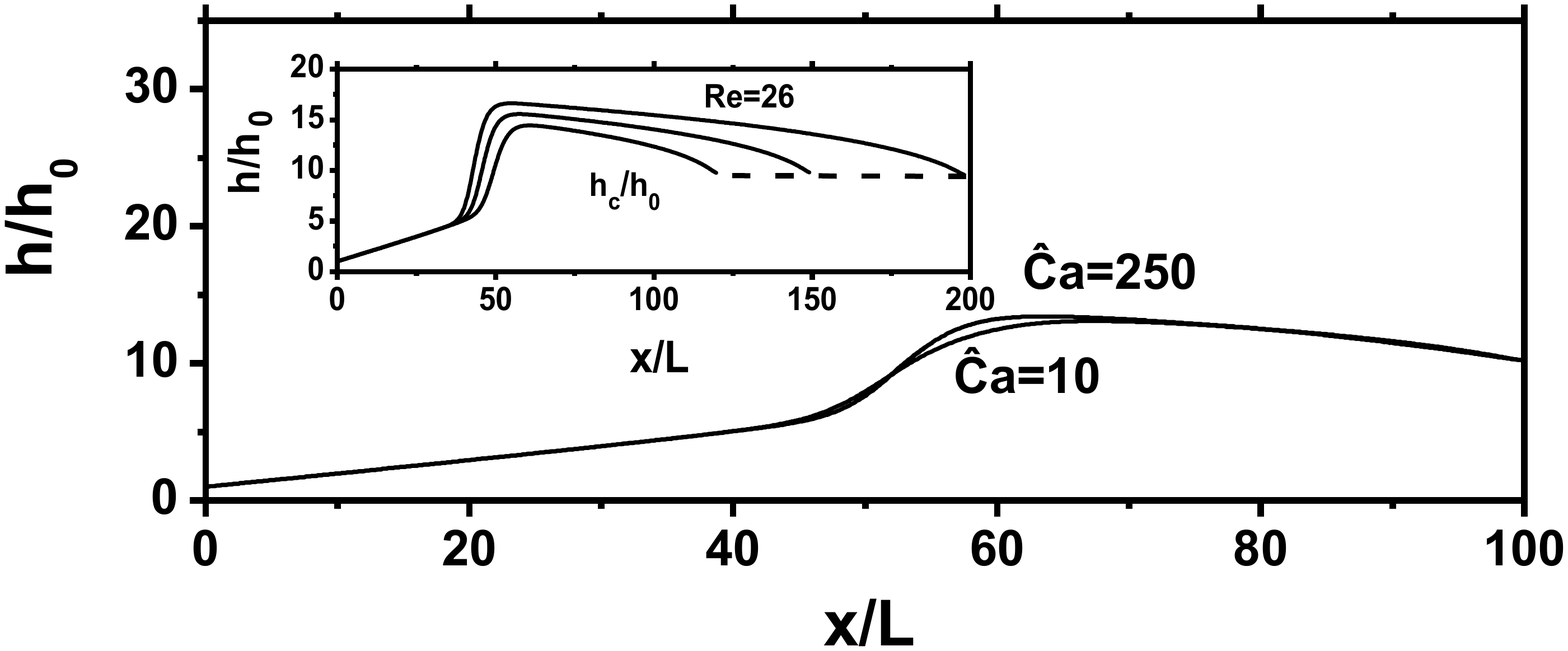}
\end{center}
\caption{Free surface profiles calculated numerically at $\theta=0^{\circ}$, $Re=26$ and $Ka=0.0013$ at different values of $\hat{Ca}$. The insert shows simulations with all parameters fixed at $\theta=0^{\circ}$,  $Re = 26, \hat{Ca} = 108, Ka= 0.0013$ , but in different domains.} 
\label{Fig11}
\end{figure}

\section*{Results and discussion}

The third order partial differential equation has sufficient degrees of freedom to satisfy boundary conditions at the inlet and the outlet of the flow. We solve problem (\ref{LE3})-(\ref{GE2D3sOne}) numerically with two boundary conditions at each end of the interval $x\in [0,l_s]$ to reach a steady state in time. That is, 
\begin{equation}
\label{Inlet}
h(0,t)=1, \quad q(0,t)=1
\end{equation}
at the inlet at $x=0$, and
\begin{equation}
\label{Outlet}
h(l_s,t)=h_c, \quad q(l_s,t)=1
\end{equation}
at the outlet at $x=l_s$.

The simulation results are illustrated in Figs. \ref{Fig41}-\ref{Fig11} in different scenarios and at variations of different parameters. 

\subsection*{Laminar flow regime}

First of all, we obtained a reference point by simulating free surface profiles in the parameter range relevant to the laminar experiments in water with a planar hydraulic jump developed~\cite{Dhar2020}. We have numerically generated a profile observed at $\theta=0^{\circ}$ and $q_0=1.25\cdot10^{-4}\,\mbox{m}^2/\mbox{s}$, which corresponds to non-dimensional parameters $Re=12.5$, $\hat{Ca}=6.9$ and $Ka=0.0098$, Fig. \ref{Fig41}. The experimental profile has a pronounced linear part in the upstream region and a hydraulic jump followed by a smooth downstream profile. 

As one can observe, Fig. \ref{Fig41}, the numerical solution provides a good approximation to the experimental dependence within the experimental error involved. It closely follows the asymptotic solution in the upstream region till the point where the jump starts developing. In the downstream region, the numerical solution follows the asymptotic solution straight after the maximum is achieved, as is expected.  

Remarkably, the inflection points of the experimental and numerical profiles, $x_E$ and $x_N$, corresponding to the midpoints of the jump region were found to be very close to each other.

In the second set of simulations, the non-dimensional parameters were scaled from the reference values $R^{(0)}$, $\hat{C}^{(0)}$, $Ka^{(0)}$ to mimic the change of the flow rate $q_0$ in the experiments according to $Re = Sc\, R^{(0)}$, $\hat{Ca} = Sc\, \hat{C}^{(0)}$, $Ka = Ka^{(0)}/Sc^2$, where $Sc$ was the scaling factor. One can observe the expected trends that the jump position follows $X\propto q_0^{5/3}$, Fig. \ref{Fig81}. Variations of the inclination angles in the numerical solutions also demonstrated the trends observed in the experiments, Fig. \ref{Fig81}. 

One can conclude that the full DAM provides an adequate description of laminar flows generating regular solutions with smooth, continuous free surface profiles. We note, even though the model parameters in the simulations were taken at real surface tension values for water, there is no transient behaviour observed, but a steady-state. So the transient features observed in the Navier-Stokes laminar flow simulations~\cite{Benilov2019} were either averaged out in the DAM or could be computational artefacts of the numerical scheme. 

\subsection*{Parametric dependencies of the full DAM}

Consider now how variations of the main parameters of the system affect the flow and the formation of the jump transition region. As a reference point, for a demonstration, we choose $\theta=0^{\circ}$ and larger values of the Reynolds and capillary numbers, $Re=26$, $\hat{Ca}=108$ and $Ka=0.0013$, relevant to the experimental conditions in~\cite{Bonn2009}. Variations of the surface profiles corresponding to variations of the flow rate $q_0$ are shown in Fig. \ref{Fig7}. The observed trend is expected, that is the jump regions moves away from the entry point with increasing flux rate, that is, in fact, with increasing the Reynolds number. One can also observe the effect of the far-field conditions by changing the size of the flow domain but keeping all other parameters fixed, as is demonstrated in Fig. \ref{Fig11}, inset. The larger domains provide stronger resistance to the flow so that the jump regions moves closer to the entry point. It is worth noting that the far-field conditions do not affect the upstream profiles, but the transition point. So that the local criteria are indeed insensitive to the far-field conditions. 

At the same time, the effect of the capillary numbers is very weak and practically negligible, Fig. \ref{Fig11}. That is variations of the surface tension do not have any effect on the flow itself and the developing jump region.

The observed trends are typical, that is they have been present in the numerical solutions at different parameters within the parameter range in this study.  
\bigskip

\subsection*{Turbulent flow regime}

To investigate the full DAM solutions in the parameter range relevant to the experiments in ~\cite{Bonn2009}, we have adjusted the liquid (water) viscosity to much larger values, assuming constant eddy viscosity at $\mu_{\epsilon} = 4.2\cdot 10^{-3}\,\mbox{Pa}\cdot\mbox{s}$ in the liquid. 

Taking this value of the effective viscosity as the reference point in the entire simulation domain, we adjusted the non-dimensional parameters of the model accordingly $Re=26$, $\hat{Ca}=108$ and $Ka=0.0013$. The result of the simulations is shown in Fig. \ref{Fig6}. As one can observe, the continuous steady-state solution to the system (\ref{LE3})-(\ref{GE2D3sOne}) with a hydraulic jump developed follows very closely the asymptotic solutions expected at these parameters away from the jump region. Also, surprisingly, the position of the jump follows the experimental observations, not just at the parameter values used in the comparison, Fig. \ref{Fig6}, but in the entire parameter range. We do not have a clear explanation for this effect at the moment, and it requires further studies, possibly using the full Navier-Stokes model with sufficient resolution to simulate small scale eddy motion.  

At the same time, the numerical solution goes below the observed free surface profile in the downstream region. This shows the limitation of the DAM approach, on the one hand, on the other hand, the result demonstrates that the effects of the developed eddy motion are by far more important (or at least can match) than the contribution from the far-field conditions or the weak surface tension effects. 

The observed deviation also demonstrates that a simple averaging approach to obtain any practical criteria for the developing hydraulic jump instability could easily fail and mislead in the presence of strong eddy motion in the flow. One effect that needs further studies is why the position of the jump in the DAM solution was so close to the experimental values. 

\section*{Conclusions}

We have shown that the full DAM possesses continuous, regular solutions manifesting instability leading to the formation of a steady hydraulic jump, and thus continuously linking the two regions, the upstream and the downstream of the jump region. A comparison with experimental data has shown that the numerical solution can adequately describe flows with hydraulic jumps in a laminar flow regime.

The obtained numerical solutions demonstrate anticipated trends with variations of the flux density, and the non-dimensional parameters of the problem. At the same time, the effect of the surface tension is found to be negligible in the formation of a planar hydraulic jump in the laminar flow regime.

The results of our analysis demonstrate that the full DAM can be used for experimental data analysis and as a benchmark case in the laminar flow regimes.

A comparison with the data where flow turbulisation took place, on the other hand, shows that simple re-normalisation of viscosity, in this case, is insufficient to adequately describe the flow and the formation of the jump region. So that, in general, the problem requires adequate tools to include short-scale eddy motion, which is dominant at high Reynolds numbers. 

\begin{acknowledgments}
The authors are grateful to Prof. D. Ian Wilson and Dr. Rajesh Bhagat for useful discussions. EC was supported through a PhD scholarship awarded by the “EPSRC Centre for Doctoral Training in the Mathematics of Planet Earth at Imperial College London and the University of Reading” EP/L016613/1.
\end{acknowledgments}

\bigskip
\begin{table*}[ht!]
\resizebox{1.7\columnwidth}{!}{
    \begin{tabular}{ | c | c | c | c | c | c | c |}
      \hline 
      Set & Liquid/Regime & $\mu$ ($\,\mbox{mPa}\cdot\mbox{s}$) at $20^{\circ} C$  & $\gamma$ (\mbox{mN/m}) & $h_0$ (mm) & $l_s$ (m) & $d$ (cm) \\  
      \hline
			
	     I: Singha et al. 2005~\cite{Ray2005}	&	Water/Turbulent & $1.0$  & $72.8$ & $\approx 0.5$ & $0.7$  & $9$ \\  
      \hline
			
			II: Bonn et al. 2009~\cite{Bonn2009}	&	Water/Turbulent & $1.0$  & $72.8$ & $\approx 0.5$ & $0.5$  & $0.8$ \\  
      \hline

			III: Bhagat	et al. 2017~\cite{Ian2017} &	Water/Turbulent & $1.0$  & $72.8$ & $0.85$ & $2.5$   & $15$ \\  
      \hline
																
			IV: Bhagat et al.	2017~\cite{Ian2017} &	WP (95/5) \footnote{5\% (w/w) 1-propanol in water} /Turbulent & $1.26$  & $42.5$ & $0.85$ & $2.5$  & $15$ \\ 
      \hline			
			
			V: Dhar et al.	2020~\cite{Dhar2020} &	Water/Laminar  & $1.0$  & $72.8$ & $0.25$ & $0.09$  & $10$ \\ 			
      \hline

    \end{tabular} }
    \caption{Parameters of the experiments with a planar hydraulic jump observed: dynamic viscosity $\mu$, surface tension $\gamma$, initial film thickness $h_0$, the total length of the setup up to the far-field $l_s$ and the channel width $d$.}
    \vspace{5pt}
    \label{Table1}
\end{table*}

\appendix*

\section{The approximation of thin films}

If we start from the full system of the Navier-Stokes equations for Newtonian liquids, introduce non-dimensional variables, that is the coordinates $x_1=\hat{x}_1/L$, $x_2=\hat{x}_2/L$, $x_3=\hat{x}_3/H$, velocities $v_1=\hat{v}_1/U$, $v_2=\hat{v}_2/U$, $v_3=\hat{v}_3/\delta U$, time $t/t_0$ and pressure $p=\hat{p}/p_0$, neglect terms of the order of $O(\delta^2)$, but assuming that the Reynolds number $Re=\delta \frac{\rho U H}{\mu} \sim O(1)$, one arrives at 
\begin{equation}
\label{LE1}
\pdiff{v_k}{x_k}=0
\end{equation}

\begin{equation}
\label{Prandtl-I}
Re \left\{ \pdiff{v_1}{t}+v_l\pdiff{v_1}{x_l}\right\}=- Re \pdiff{p}{x_1} + 
\end{equation}
$$
\pdiff[2]{v_1}{x_3}+  Re\, Ka_{||}\, \sin\theta,
$$

\begin{equation}
\label{Prandtl-II}
 Re \left\{ \pdiff{v_2}{t}+v_l\pdiff{v_2}{x_l}\right\}=-  Re \pdiff{p}{x_2} + \pdiff[2]{v_2}{x_3}, 
\end{equation}

\begin{equation}
\label{Prandtl-III}
0=\displaystyle \pdiff{p}{x_3} +  Ka\, \cos\theta.
\end{equation}
Here $L$ and $H$ are the vertical and horizontal length scales respectively, $\delta=H/L\ll 1$ is the small parameter of the problem, $U$ is characteristic velocity, $t_0=L/U$ is the timescale, $p_0=\rho U^2$ is the characteristic pressure in the inertial range, $\rho$ is the liquid density and $\mu$ is dynamic viscosity. The non-dimensional parameters featured in the formulation are (in addition to the Reynolds number $Re$) $Ka_{||} =\frac{g_0 L}{U^2}$, $Ka=\frac{g_0 H}{U^2}$.

The reduced system of the Navier-Stokes equations (\ref{LE1})-(\ref{Prandtl-III}) should be augmented with the boundary conditions on the solid at $x_3=B(x_1,x_2)$ and the free surface at $x_3=h(x_1,x_2,t) + B(x_1,x_2)$. Keeping leading order terms at $\delta\ll 1$, we have no-slip and impermeability boundary conditions at  $x_3=B(x_1,x_2)$
\begin{equation}
\label{nosliptf}
v_k=0, \quad k=1,2,3,
\end{equation}
the zero stress, the normal stress and the kinematic boundary conditions at the free surface $x_3=h(x_1,x_2,t) + B(x_1,x_2)$
\begin{equation}
\label{zerosttf}
\pdiff{v_1}{x_3}=0, \quad \pdiff{v_2}{x_3}=0,
\end{equation}
\begin{equation}
\label{nstf}
p=p_a-\frac{1}{\hat{Ca}\, Re}\left(\frac{\partial^2 (h+B)}{\partial x_1^2} + \frac{\partial^2 (h+B)}{\partial x_2^2} \right)
\end{equation} 
and
\begin{equation}
\label{kctf}
v_3=\pdiff{h}{t} + v_1\pdiff{(h+B)}{x_1} + v_2\pdiff{(h+B)}{x_2}.
\end{equation}
Here $p_a$ is external gas pressure and $\hat{Ca} = Ca\, \delta^{-3}$ is a renormalised capillary number $Ca=\frac{\mu U}{\gamma}$, $\gamma$ is surface tension of the liquid. 

Using (\ref{Prandtl-III}) and boundary condition (\ref{nstf}), one can resolve pressure explicitly
\begin{equation}
\label{PrTF}
p= p_a+Ka\, \cos\theta\, (h+B-x_3) -
\end{equation}
$$
\frac{1}{\hat{Ca}\, Re}\left(\frac{\partial^2 (h+B)}{\partial x_1^2}   +  \frac{\partial^2 (h+B)}{\partial x_2^2}\right). 
$$
 
\subsubsection*{Karman-Pohlhausen approach to averaged equations}

Integrating the incompressibility condition (\ref{LE1}) from $x_3=B$ to $x_3=h+B$, using (\ref{kctf}) one gets
\begin{equation}
\label{LE2}
\frac{\partial h}{\partial t}  + \frac{\partial q_1}{\partial x_1} + \frac{\partial q_2}{\partial x_2} = 0,
\end{equation}
where $q_{1,2}=\int_{B}^{h+B}\,v_{1,2}\, dx_3$.

Integrating the remaining Navier-Stokes equations (\ref{Prandtl-I})-(\ref{Prandtl-II})
\begin{equation}
\label{Prandtl-ISLTT}
 \frac{\partial q_1}{\partial t} + \frac{\partial }{\partial x_1} \int_{B}^{h+B} v_1 v_1 dx_3 +  \frac{\partial }{\partial x_2} \int_{B}^{h+B} v_1 v_2 dx_3=
\end{equation}
$$
h\, \left\{ Ka_{||}\sin\theta -   \pdiff{p}{x_1} \right\} -\frac{1}{Re} \left. \pdiff{v_1}{x_3}\right|_{x_3=B},
$$

\begin{equation}
\label{Prandtl-IISLTT}
 \frac{\partial q_2}{\partial t} +  \frac{\partial }{\partial x_1} \int_{B}^{h+B} v_2 v_1 dx_3 +  \frac{\partial }{\partial x_2} \int_{B}^{h+B} v_2 v_2 dx_3=
\end{equation}
$$
-h\,  \pdiff{p}{x_2}  -\frac{1}{Re}  \left. \pdiff{v_2}{x_3}\right|_{x_3=B}.
$$

Using the Karman-Pohlhausen ansatz
\begin{equation}
\label{ansatz}
v_{1,2}
=-\frac{3}{2}\frac{q_{1,2}}{h^3}\left\{ x_3^2 - 2(h+B)x_3 +B(B+2h) \right\},
\end{equation}
which satisfies no-slip condition (\ref{nosliptf}) and zero-stress condition (\ref{zerosttf}),
one finally gets
\begin{equation}
\label{GE2D3s}
\frac{\partial q_1}{\partial t} + \frac{6}{5} \frac{\partial }{\partial x_1} \left( \frac{q_1^2}{h} \right) +  \frac{6}{5} \frac{\partial }{\partial x_2} \left( \frac{q_1q_2}{h} \right) = 
\end{equation}
$$
h\, \left\{ Ka_{||}\sin\theta -   \pdiff{p}{x_1} \right\} -\frac{3}{Re}\frac{q_{1}}{h^2}
$$
and
\begin{equation}
\label{GE2D4s}
 \frac{\partial q_2}{\partial t} + \frac{6}{5}\frac{\partial }{\partial x_1} \left( \frac{q_1q_2}{h}\right) +  \frac{6}{5} \frac{\partial }{\partial x_2} \left( \frac{q_2^2}{h}\right)=
\end{equation}

$$
-h\,  \pdiff{p}{x_2}  -\frac{3}{Re}\frac{q_{2}}{h^2},
$$
where pressure $p$ is given by (\ref{PrTF}). System (\ref{LE2})-(\ref{GE2D4s}) is the required system in the thin film approximation.

Note, we used the following properties of the Karman-Pohlhausen ansatz for $i,j=1,2$
$$
\int_{B}^{h+B} v_{i}\,dx_3 = q_{i},
$$
$$
\left. \pdiff{v_{i}}{x_3}\right|_{x_3=B} = 3\frac{q_{i}}{h^2}
$$
and 
$$
\int_{B}^{h+B}v_i v_j\, dx_3 = \frac{9}{4}\frac{q_i q_j}{h^6}\int_{B}^{h+B} (x_3^2 -
$$
\begin{equation}
\label{KPA}
 2(h+B)x_3 +B(B+2h))^2\, dx_3 =  \frac {6q_i q_j} {5h}.
\end{equation}

\end{document}